# Title: Geospatial Big Data Handling Theory and Methods: A Review and Research Challenges

**Authors (with equal contribution):**

Songnian Li *, Ryerson University, Toronto, Canada, snli@ryerson.ca

Suzana Dragicevic, Simon Fraser University, Vancouver, Canada, suzanad@sfu.ca

François Anton, Technical University of Denmark, Lyngby, Denmark, fa@space.dtu.dk

Monika Sester, Leibniz University Hannover, Germany, monika.sester@ikg.uni-hannover.de

Stephan Winter, University of Melbourne, Australia, winter@unimelb.edu.au

Arzu Coltekin, University of Zurich, Switzerland, arzu@geo.uzh.ch

Chris Pettit, University of Melbourne, Australia, cpettit@unimelb.edu.au

Bin Jiang, University of Gävle, Sweden, bin.jiang@hig.se

James Haworth, University College London, UK, j.haworth@ucl.ac.uk

Alfred Stein, University of Twente, The Netherlands, stein@itc.nl

Tao Cheng, University College London, UK, tao.cheng@ucl.ac.uk

* Corresponding author

# Geospatial Big Data Handling Theory and Methods: A Review and Research Challenges

Abstract: Big data has now become a strong focus of global interest that is increasingly attracting the attention of academia, industry, government and other organizations. Big data can be classified in the disciplinary area of traditional geospatial data handling theory and methods. The increasing volume and varying format of collected geospatial big data presents challenges in storing, managing, processing, analysing, visualising and verifying the quality of data. This has implications for the quality of decisions made with big data. Consequently, this position paper of the International Society for Photogrammetry and Remote Sensing (ISPRS) Technical Commission II (TC II) revisits the existing geospatial data handling methods and theories to determine if they are still capable of handling emerging geospatial big data. Further, the paper synthesises problems, major issues and challenges with current developments as well as recommending what needs to be developed further in the near future.

## 1. Introduction

Over the last decade, big data has become a strong focus of global interest, increasingly attracting the attention of academia, industry, government and other organizations. The term "big data" first appeared in the scientific communities in the mid-1990s, gradually became popular around 2008 and started to be recognized in 2010. Today, big data is a buzzword everywhere on the Internet, in the trade and scientific publications and during all kinds of conferences. Big data has been suggested as a predominant source of innovation, competition and productivity (Manyika et al. 2011). The rapid growing flood of big data, coming from the many different types of sensors, messaging systems and social networks in addition to more traditional measurement and observation systems, have already invaded many aspects of our everyday existence. On the one hand, big data, including geospatial big data, has great potential to benefit many societal applications such as climate change, disease surveillance, disaster response, monitoring critical infrastructures, transportation and so on. On the other hand, big data's benefits to society are usually limited by issues such as data privacy, confidentiality and security.

Big data is still not a clearly defined term and it has been defined differently from technological, industrial, research or academic perspectives (Chen et al. 2014). In general, it is considered as structured and unstructured datasets with massive data volumes that cannot be easily captured, stored, manipulated, analysed, managed and presented by traditional hardware, software and database technologies. Along with its definitions, big data is often described by its unique characteristics. In discussing application delivery strategies under increasing data volumes, Laney (2001) first proposed three dimensions that characterise the challenges and opportunities of increasing large data volumes: *Volume*, *Velocity* and *Variety* (3Vs). While *3Vs* have been continuously used to describe big data, the additional dimension of *Veracity* has been added to describe data integrity and quality. Further *Vs* have also been suggested such as variability, validity, volatility, visibility, value or visualization. However, these are met critically as they do not necessarily express qualities of magnitude. While it is true that these further *Vs* do not directly contribute to understanding what the word big refers to in big data, they do touch on important concepts related to the entire pipeline of big data collection, processing and presentation. Suthaharan (2014) even argued that 3Vs cannot support early detection of big data characteristics for its classification and proposed 3Cs: *cardinality*, *continuity*, and *complexity*. It is apparent that defining big data and its characteristics will be an ongoing endeavour, but it nevertheless will not have negative impact on big data handling and processing.

According to the arguable phrase "80% of data is geographic" (see discussions in Morais (2012)), much of the data in the world can be geo-referenced, which indicates the importance of geospatial big data handling. Geospatial data describe objects and things with relation to geographic space, often with location coordinates in a spatial referencing system. Geospatial data are traditionally collected using ground surveying, photogrammetry and remote sensing, and more recently through laser scanning, mobile mapping, geo-located sensors, geo-tagged web contents, volunteer geographic information (VGI), global navigation satellite system (GNSS) tracking and so on. Adopting the widely accepted characterisation method, geospatial data can exhibit at least one of the *3Vs* (Evans et al. 2014), but the other *Vs* mentioned above are also relevant. As such, geospatial big data can be characterised by the following, with the first four being more fundamental and important:

- Volume: Petabyte archives for remotely sensed imagery data, ever increasing volume of real time sensor observations and location-based social media data, vast amount of VGI data, etc., as well as continuous increase of these data, raise not only data storage issues but also a massive analysis issue (Dasgupta 2013).
- Variety: map data, imagery data, geotagged text data, structured and unstructured data, raster and vector data, all these different types of data – many with complex structures – calls for more efficient models, structures, indexes and data management strategies and technologies, e.g., use of NoSQL.
- Velocity: imagery data with frequent revisits at high resolution, continuous streaming of sensor observations, Internet of Things (IoT), real-time GNSS trajectory and social media data all require matching the speed of data generation and the speed of data processing to meet demand (Dasgupta 2013).
- Veracity: much of geospatial big data are from unverified sources with low or unknown accuracy, level of accuracy varies depending on data sources, raising issues on quality assessment of source data and how to "statistically" improve the quality of analysis results.
- Visualization: provides valuable procedures to impose human thinking into big data analysis. Visualizations help analysts identifying patterns (such as outliers and clusters), leading to new hypotheses as well as efficient ways to partition the data for further computational analysis. Visualizations also help end users to better grasp and communicate dominant patterns and relationships that emerge from the big data analysis.
- Visibility: the emergence of cloud computing and cloud storage has made it possible to now efficiently access and process geospatial big data in ways that were not previously possible. Cloud technology is still evolving and once issues such as data provenance – historical metadata – are resolved, big data and the cloud would be mutually dependent and reinforcing technologies.

The increasing volume and varying format of collected geospatial big data pose additional challenges in storing, managing, processing, analysing, visualising and verifying the quality of data. Shekhar, et al. (2012, p. 1) states that "the size, variety and update rate of datasets exceed the capacity of commonly used spatial computing and spatial database technologies to learn, manage, and process the data with reasonable effort". Big data tends to hold people to expect more and larger hypotheses that grow faster than the statistical strength of data and capacity of data analysis (Gomes 2014). Verifying the quality of geospatial big data and data products delivered to end users is noted as one of the big challenges and becomes even more challenging in the quality control of the delivered data products (see 2012 ISPRS Resolution, www.isprs.org/documents/resolutions.aspx). On the other hand, fitness of uses or purposes appears more valid or should be advocated (Mayer-Schönberger & Cukier 2013) in the context of big data.

The objectives of this paper are to 1) revisit the existing geospatial data handling methods and theories to determine if they are still capable of handling emerging geospatial big data; 2) examine current, state-of-the-art methodological, theoretical, and technical developments in modelling, processing, analysing and visualising geospatial big data; 3) synthesize problems, major issues and challenges in current developments; and 4) recommend what needs to be developed in the near future. Section 2 to 6 addresses objectives 1 and 2 of the 5 important areas related to geospatial big data handling methods and theories, which are the focus of various Working Groups (WG) of ISPRS TC II. Related image analysis and processing topics, such as dimensionality reduction; image compression; compressive sensing in big data analytics; content-based image retrieval; and image endmember extraction, are not covered in this paper. Section 7 presents open issues and future research directions of the three focus areas of TC II. Section 8 gives a summary and conclusions to the paper.

## 2. Collection of Geospatial Big Data

In recent years, along with the availability of new sensors, new ways of collecting geospatial data have emerged, leading to completely new data sources and data types of geographical nature. Data acquired by the public, so-called Volunteered Geographic Information (VGI), and data from geo-sensor networks have led to an increased availability of spatial information. Whereas until recently, authoritative datasets were dominating in topographic domain, these new data types extend and enrich geographic data in terms of thematic variation and by the fact that it is more user-centric. The latter is especially true for VGI collected by social media (Sester et al. 2014).

Geospatial data collection is shifting from a data sparse to a data rich paradigm. Whereas some years back geospatial data capture was based on technically demanding, accurate, expensive and complicated devices, where the measurement process was itself sometimes an art, we are now facing a situation where geospatial data acquisition is a commodity implemented in everyday devices used by many people. The devices are capable of acquiring environmental geospatial information at an unprecedented level, with respect to geometric and temporal resolution and thematic granularity. They are small, easy to handle, and able to acquire data even unconsciously.

This data capture paradigm is similar to the situation in topographic data collection for digital terrain models by capturing significant topographic points with morphological characteristics on the one hand ("qualified" points, i.e., points with semantics)– as opposed to the collection of point clouds using LiDAR sensors or stereo matching, leading to masses of "unqualified" points (Ackermann 1994). The first approach requires manual selection and measurement and guarantees that the topographic reality can be interpolated from the sparse measurements. The second approach assumes that the topographic reality is captured by the dense measurements and can be reconstructed from them – thus the object formation and identification are shifted to the analysis process.

In general, one can distinguish the following sensor configurations: 1) objects equipped with sensors moving through space and capturing their own trajectories /and the local environments: humans and moving devices such as cars; and 2) static sensors constantly observing the (changing) environment. Today's data acquired by these new sensors and new stakeholders can be characterised as follows:

- data streams
- arbitrary high density
- "close sensing" (Duckham 2013), i.e., the ability to measure many different dimensions of objects characteristics, e.g., optical, acoustic, and mechanical features
- different degrees of positional accuracy and reference, ranging from highly precise coordinates via relative positions to information in which has no geometric reference or is only implicitly located by location names

Ample example data collections exist, which may lead to geospatial big data sets. From a "social" perspective, for example, over the last decade we have seen (through the rise of the so-called "smart city" concept) the instrumentation of cities which are now providing vast amount of real-time data through the likes of smart card ticketing systems, vehicle tracking devices, CCTV, toll systems, induction loops and other sensors. With the rise of social media we are also seeing vast amounts of data (e.g., from Twitter feeds), which can be geotagged and used to assist in disaster management and emergency relief. From an "environmental science" perspective, there are huge remotely sensed imagery repositories such as NASA's Landsat repositories which provide petabytes of geospatial data (Riebeek 2015). Capturing the urban environment is possible with a variety of sensors today. Cars (e.g., connected vehicles) are equipped with a lot of sensors in order to assist the driver and enhance his/her safety and comfort, many of them also capturing the immediate environment of the car, e.g., front cameras, backwards cameras, ultrasonic (for parking assistance), GPS, radar, rain-sensing wipers (Fitzner et al. 2013). The information is stored on the local bus system, but can also be transmitted to the surrounding infrastructure or to other vehicles.

Another important example concerns the quality of data on health. Such data are routinely collected and stored, e.g., with doctors or health centres. In particular in public health centres, however, the coordinates in space and time may lack quality. The first reason is that health aspects are not always related to the location where the person lives. The second reason is that the moment when he or she is visiting the health facility may not correspond with the time of incidence. As an important reason we found that people avoid stigmatic investigations, e.g., related to AIDS or other sexually transmitted diseases, thus giving a bias in routinely collected datasets (Kandwal et al. 2010).

People, considered as "sensors", can also help capture traffic or mobility related VGI style information. VGI acquisition can be distinguished into participatory and opportunistic. Participatory data acquisition is conducted in a conscious process by a user, who explicitly selects objects and their features and contributes this information (an example is the OpenStreetMap, OSM). Opportunistic data capture occurs unconsciously, mostly with no specific purpose – or even a completely different purpose. A prominent example is the exploitation of mobile phone data to determine traffic information such as traffic jams. The capture of the spatio-temporal phenomenon "traffic jam" is just a by-product of many users having switched their phones on when driving their cars. With the recent emergence of smart cards, transport ticketing systems like the London Oyster card are capturing the movement of millions of travellers which use the London Tube and railway system daily. A new data point is created every time they register the location via swiping on or off a mode of public transport.

VGI has proven to be an essential data source, when ad-hoc information is needed as in a disaster context (Goodchild & Glennon 2010). In those situations, it is important to get any information – even if it is not very accurate. Thus, social media and services are considered as new information sources for example for early response and crisis management (Fuchs et al. 2013). Fuchs et al. (2013) evaluated Twitter streams to detect large scale flooding events in Germany. In a period of 8 months, approximately 6 million tweets had been recorded. If the analysis concentrates only on the frequency, it was not possible to identify the events; however, the inclusion of specific keywords, together with spatio-temporal clustering was able to detect some of the events. A similar approach is reported by Dittrich and Lucas (2013). Huang et al. (2015) used millions of location-based tweets to predict human movements. Other examples for the successful use of crowd-sourcing data collection approach in the context of disasters, are the Haiti earthquake (http://www.ushahidi.com/), the Queensland flood (McDougall 2011), as well as flood risk assessment (Poser et al. 2009). Invitation active participation of users in the context of mapping has been reported by Frommberger et al. (2013).

It is worth noting that in the realm of geoscientific data, there is a wealth of new sensors and data sources, which lead to large collections of diverse and "dirty" data (inaccurate, incomplete or erroneous data), which only gain relevance by careful integration and fusion with complementary data (Van Zyl et al. 2009).

## 3. Quality Assessment

Geospatial data, being abstractions and observations of a continuous reality (Frank 2001), is by nature uncertain, ideally time-stamped and often incomplete. Accordingly, geospatial big data, with its defining characteristics of being large (voluminous), heterogeneous (variety), real-time processed (velocity), inconsistent (variability), and thus also of variable quality (veracity), must suffer even more from uncertainty, asynchronicity, and incompleteness. However, while certain effects on data quality are emphasized for geospatial big data, the phenomena to be described are still the same. Thus, the known methods and theories of quality assessment are still applicable.

In geographic information science and technology, standardized methods have been developed in order to assess, describe and propagate quality characteristics both quantitatively and qualitatively. Frameworks exist to describe data quality from a producer's perspective, which then, in the hands of consumers, have to be translated into fitness for purpose. These frameworks went into international standards, such as ISO 8402 (which is generically about quality management: "Data quality is the totality of characteristics of a product that bear on its ability to satisfy stated and implied needs") or ISO 19157 (which is specifically about the quality of geographic information: "Establishes the principles for describing the quality of geographic data. It defines components for describing data quality, specifies components and content structure of a register for data quality measures, describes general procedures for evaluating the quality of geographic data, and establishes principles for reporting data quality"). The frameworks typically define quantitative measures of data quality, such as spatial, temporal and thematic accuracy, spatial, temporal and thematic resolution, consistency, and completeness (Veregin 2005), and in addition qualitative characteristics of data quality, such as purpose, usage, or lineage.

In the end, this approach to data quality assessment is descriptive about the data capturing process, stored in separate metadata. But these data are used in decision making processes, and thus, the quantities and qualities of the above frameworks require further analysis for propagation. In a spatial statistical context work has been done by Van de Vlag et al. (2005) and Van de Vlag and Stein (2006) on natural objects and Kohli et al. (2012)

on slums. Frank (2007, p. 417), studying the ontology of imperfect knowledge, stated: "How do the imperfections in the data affect the decision?", but acknowledges that the decision making process is a black box of unknown complexity. While still some quality descriptors, especially the quantitative ones, lend themselves to functional error propagation, others – especially the qualitative ones – are still left to human judgment. With the ad-hoc combination of data streams in geospatial big data collection and near-real-time analytics, this traditional approach falls short both in collecting, aggregating or propagating metadata as well as in human judgment of metadata. These challenges can be illustrated for example in the context of connected urban transport introduced above:

- The sheer volume of geospatial big data in transport arises from the large number of agents (vehicles, people, and goods) on their way at any time. Furthermore, in order to be able to predict transport demand or traffic, not only are real-time data required but also historic data. If the bandwidth of communication channels forms bottlenecks, either the sampling rate or the sampling size can be reduced, or the computation can be decentralized (Duckham 2013), in which case a central instance would collect only aggregated data. Each of these solutions has an impact on the quality assessment of the collected data.
- The large variety of geospatial big data in transport (such as GNSS, inertial sensors, compass, wheel sensors, radar, laser scanning, number plate recognition, induction loops, electronic toll or ticketing, parking sensors, social media comments, citizen reports, to name only a few) to be combined in big data analytics challenges error propagation models, which are based on a functional relationship.
- The velocity of data, more often than not requiring real-time analysis – including real-time integration – for event management or near-future prediction, does not allow for setting up proper error propagation.
- The variability of geospatial big data in transport, indicating inconsistency, stems from both the variety between different data channels as well as the unreliability of any of these sources. For example, volunteered geographic information can be inconsistent if the "citizens as sensors" (Goodchild 2007) or "citizens as databases" (Richter & Winter 2011) disagree on their observations, are just not sampling a particular phenomenon of interest (e.g., if a tree falls on a street and nobody is there to notice), or are handicapped by lack of communication channels (e.g., a sensor is moving out of reach of WiFi/cell phone coverage). In another example, satellite positioning has some well-known quality descriptors, but in urban canyons, these measures vary with the location in the environment rather than the sensor, the transmission channel or the time (Kealy et al. 2014). This dependency on location is non-linear and difficult to predict.
- The veracity of geospatial big data in transport indicates already that data is to be combined of very different quality. This extends also to irregular sampling rates (both spatial and temporal), entry errors, redundancy, corruption, lack of synchronization, or a variety of collection purposes (taxonomies, semantics), to name a few.
- Variability and veracity are closely tied to vulnerability, which is perhaps the only aspect in contrast to classic institutional databases, which are behind firewalls or even completely disconnected (in safety critical applications). Since both big data collection and analytics require connectivity, concepts are also required to deal with malicious contributions, attacks and theft, and privacy. The latter is particularly true for geospatial data collected for tracking movements, and a measure of quality would cater for protection of privacy while still guaranteeing some level of quality of service (Anthony et al. 2007, Duckham & Kulik 2006).

Within geospatial big data in transport, a prominent example is research watching the quality of OpenStreetMap, which has become an open source for navigation services at a global scale. This research focuses mostly on the completeness of OpenStreetMap (Haklay 2010, Mondzech & Sester 2011, Neis et al. 2012, Zielstra & Zipf 2010).

Typically, the value of geospatial big data (analytics) is seen in information for decision support. Analytical methods such as data mining and machine learning enable only inductive reasoning on big data, i.e., detection of global correlations, or predictions based on these correlations. In transport, an example is the early discovery of traffic accidents. In these applications the traditional quality (from a provider or consumer perspective) has been replaced by correlation coefficients (Miller & Goodchild 2014), well knowing that correlation is not necessarily about causes or truth. Thus, validity or trust is traded for the velocity of information production.

# 4. Data Modelling and Structuring

All the spatial data models, including the spaghetti vector data model, the network data model, the topological data model and the regular (raster) and irregular (Voronoi, k-d-tree, Binary Space Partitioning Tree, paving group, crystallographic group) tessellations based data models, can be used for handling geospatial big data. Nevertheless, there are some models that are more suitable for handling very large data sets and others that are less suitable for geospatial big data. As the network and topological data models need to store the connectivity (for the network and topological spatial data model) and the adjacency (for the topological spatial data model), they are not well suited for handling geospatial big data streams unless a very efficient spatial data indexing is making the update of the connectivity and/or the topology possible in real time. Since geospatial big data environments might have to relax accuracy constraints to satisfy the real-time constraint, irregular and especially regular tessellations are ideally suited to handle soft errors - inconsistencies in data that do not necessarily cause erroneous results. Finally, as regular tessellations can be stored in matrices, they are subject to very parallelized vectorization algorithms. Nevertheless, the implicit topology of the regular tessellation (raster) spatial data model might not be desired topology in some applications that require small inaccuracies (e.g., collision detection for driver-less cars). In such cases, the only possible way to satisfy real time constraints is to use a spatial data indexing method that can maintain its performances with a big data stream being updated in real time. Current spatial data indexing methods cannot handle geospatial big data streams, because their efficiency gets lower as new spatial data streams go over the capacity of extension of the spatial data index (e.g., all the available locations for a hash function value get spent and the hash function needs to get updated or a tree must be rebalanced after a series of additions of data from the stream).

Spatial statistics is exceptionally suited to handle big data. It offers opportunities to summarize the data, and express measures of variation and uncertainty. The big concern, however, is that many of the processes and procedures are developed for the smaller datasets. In particular, much of spatial statistical analysis is either done on datasets that are collected at a pointwise scale (such as field data, meteorological data, or administrative data) and of a relatively small content, or is focusing on the relatively large image data sets which have a very specific nature. Spatial statistics depends upon the notion of spatial (and spatio-temporal) dependence, and such dependence in turn depends upon the notion of distance between points. For n observations, including their coordinates in space or space and time, evaluating distances requires inspection of $n^2$ pairs of points, and here steps should be made to be able to do this efficiently. The current data structures as such are usually able to handle the big data as well, but most likely specific procedures have to be developed that are able to address issues that are relatively novel (such as combining data in the space-time domain) or that have to address specific questions and problems, i.e., to select data from a big data set for a particular model application. A particular way ahead may be that classification of the data into multiple classes is done in the form of metadata. In such a way it is possible to make the big data of relevance in a wide range of practical applications. This would require an improved database structure, and in particular a very much adaptive spatial statistical analysis procedure.

Some authors have already pointed out the necessity of parallel and distributed programming for handling the big data sets in the general context or even in the geospatial context (Lee et al. 2014, Shekhar et al. 2012, Shekhar et al. 2014, Wang et al. 2013). Others have pointed out the usefulness of functional programming concepts or languages such as Haskell Domain-Specific Language (Mintchev 2014), Map-reduce (Maitrey & Jha 2015, Mohammed et al. 2014), Data Flow Graphs (Tran et al. 2012), or self-adjusting computation (Acar & Chen 2013). However, there is a gap between the research works that advocate functional programming techniques but do not handle specifically geospatial data, and research works that focus on geospatial big data, but do not guarantee the absence of data races (which are the races of different threads to gain access to the same data item in some shared memory [Milewski 2009]). The following sections discuss issues related to functional programming paradigms for big data streams and geospatial big data analytics in the context of big data modeling and sturcturing, and examines how geospatial data models and structures are adapted to big data.

## *4.1 Functional programming for big data streams*

The main stumbling block for handling geospatial big data streams using parallel programming and the best reason for using functional programming is the concept of data races. A data race is a race between different threads that try to access the same data items, and relates to the notion of concurrency. Functional programming solves the problem of data races by strictly controlling the simultaneous access to mutable data. It has been predicted that data races will produce the "downfall of imperative programming" (Milewski 2009). The main advantage of Haskell (or other functional programming languages like Closure, Lisp, ML, Scheme) for big data is its support of parallel computing and concurrency and the high performance of the most famous Haskell compiler (GHC: the Glasgow Haskell Compiler): "applications built with GHC enjoy solid multicore performance and can handle hundreds of thousands of concurrent network connections" (Mintchev 2014).

Domain Specific Languages (DSL) provide a solution to the key challenge of big data by making the "multi-disciplinary collaboration as effective and productive as possible" and by offering the "required degree of flexibility and control" and a development completed on time (Mintchev 2014). The functional programming languages map and reduce functions are the basis of the MapReduce programming model for processing big data sets by a distributed parallel algorithm. Maitrey & Jha (2015) states "MapReduce has emerged as the most popular computing paradigm for parallel, batch-style and analysis of large amount of data", especially since Google adopted it.

### *4.2 Geospatial big data analytics*

We can classify the techniques used for spatial data mining from different points of view: the assumptions that these techniques pre-suppose and the "curse of dimensionality" that they exhibit or not. While parametric statistics assume some probability distribution function or some spatial distribution, non-parametric statistics only assume local smoothness. Finally, functional analysis (e.g., wavelets) and homotopy continuation techniques assume only the continuity of the functions involved. While statistical and machine learning techniques exhibit the "curse of dimensionality" (Juditsky et al. 1995), homotopy continuation techniques are not subject to an exponential growth of the number of samplings and the uncertainty of the modeling does not explode from one dimension to the next one (Musiige et al. 2013, Musiige et al. 2011). This observation can be extended to other functional analysis techniques as wavelets (Juditsky et al. 1995).

As it can be conceived, in any attempt to process big spatial data streams in real time, one might be tempted to ask oneself if tolerating soft errors could be feasible, and in which extent. This was addressed in Carbin et al. (2013). If we do not accept soft errors, then we need to rely on new High Performance Computing architectures to harness the parallelism necessary to process geospatial big data streams in real time (Carbin et al. 2013). However, in order to analyze and compare geospatial big data algorithms, we need benchmarks, so that the different algorithms can have a common evaluation basis. Big data benchmarks (Shekhar et al. 2014, Shekhar et al. 2012) have become one fundamental concept in studying geospatial big data.

Spatial databases research has intended to make query processing faster by designing spatial indexing methods, which partition the search space in tiles (possibly irregular), so that the average query time will be concentrated in one tile. The main challenge is to organize geospatial data in tiles cleverly so that the access to n-dimensional data is done efficiently by referring to the location of the tile along a "space-filling curve". Several space-filling curves have been proposed in the literature. However, the Hilbert space-filling curve has the advantage that the arc length between two consecutive tiles along the space-filling curve is constant (Ujang et al. 2014). The main use and value of geospatial big data is to dig useful information from geospatial big data sets (Wang et al. 2013, Wang & Yuan 2013, 2014).

## 5. Data Visualization and Visual Analytics

When several additional Vs are proposed in defining the big data (see Section 1 above), it is no surprise that the terms *visualization* and *visual analytics* are frequently mentioned. There are various good reasons for this; primary one being that some of our computational and statistical approaches do not scale – there is simply *too much* data (Keim et al. 2013, Shneiderman 2014). Even smaller amounts of data in forms and tables are not really human readable, thus the interactive and exploratory visualization environments help at the very early stages in dealing with big data in making sense of what the data actually contains (Cook et al. 2012, Frankel & Reid 2008, Hoffer 2014). In other words, visualizations essentially enable humans to deal with big data where machines might fall short. Conversely, visual analytics approaches acknowledge the human shortcomings as well and combine the powers of computational tools with powers of human visual sense-making (Choo & Park 2013). Visualizations, therefore, are widely acknowledged as a part of analysis process (i.e., not only communication), in which we can explore the data, and build hypotheses during this process (Zhang et al. 2012). It is important to note that visualizations have been more commonly conceptualized as *communication* tools. While this is very true and visualizations are very important in communicating hypotheses, results and ideas, in the case of big data, we believe their role in *exploration* plays an even more important part. However, visualizations, while often offered as remedies to the shortcomings of the computational methods, may not scale either. Big data means many things to display; and it often results in very 'busy' displays, especially given the trend for multiple-linked view displays that are popular in visual analytics and big data applications. Such requirements can lead to information overload. It is important to note that human cognitive resources (such as the visual working memory that is critical in processing visual information; or spatial abilities which are critical for how well we can make sense of visualizations) are limited (e.g., Hegarty 2011, Hegarty et al. 2012). Novel visualization designs are necessary, especially those that are informed by knowledge on human information processing, perception and cognition.

In terms of novel design and visualization paradigms, the field witnessed many alternative approaches that are constantly being proposed, even though many of them are not yet validated through empirical testing. Among these approaches, most dominant one appears to be the *multiple-linked views*, in which approaches such as brushing and linking are utilized to allow the viewer to work with various visualizations at the same time (Bernasocchi et al. 2012). Alternatively, summarization, clustering and highlighting approaches have been proposed. Vision-inspired approaches, such as *focus+context* visualizations and *foveation* may provide interesting new opportunities as they attempt reducing the information load, and are becoming more relevant with the recent technological developments offering cheaper and better eye tracking solutions (Bektas & Çöltekin 2012, Cockburn et al. 2008, Çöltekin 2009). Other modern approaches have also been proposed in literature taking advantage of technological developments such as *cloud computing*, *parallel processing*, *indexing and querying* for real time utilization (e.g., Chen et al. 2014, Hoffman 2012, Liu et al. 2013, Lu et al. 2013).

Despite an influx of "new" solutions, we have also witnessed a rediscovery of a "not so new" system -- through a strong coupling between geographic information systems (GIS) and big data. GIS is an unmatched and mature toolbox for *data science* as its abilities to process spatial and non-spatial (attribute) data (even when they are not perfectly structured) through computational as well as visual means. Some even called GIS and big data "two parts of a whole" (Deogawanka 2014). Geographic information science and related domains such as remote sensing and geoinformatics have been dealing with large datasets for a relatively long time, before the term big data took momentum in science, popular culture and business (Çöltekin & Reichenbacher 2011). With the advent of big data, other branches of geography have also shown great interest in utilizing big data for addressing social (human geography) and environmental (physical geography) questions (e.g., Crampton et al. 2013, Goodchild 2013, Kitchin 2013, Steed et al. 2013, Wood et al. 2013).

One of the challenges is to be able to make such geospatial big data accessible to end users so that it can be used to make real world decisions. Data visualization tools and techniques are therefore critical in providing windows into such rich data so that it can be analyzed and interrogated by researchers, policy and decision-makers and citizens alike. World Wide Web platforms, such as *geoportals*, provide an excellent means to deliver such services. Portals provide access to geospatial data, and there has been significant momentum in creating federated geoportal which can access a window to a vast array of geospatial data sets. For example, the INSPIRE Geoportal (http://inspire-geoportal.ec.europa.eu/) provides access to 10,000s of geospatial metadata data records from across Europe. The INSPIRE Geoportal uses a visualization interface comprising of both a map window and folksonomy tag cloud to assist users in navigating this rich geospatial data resource.

There has been an explosion of data available and as our planet continues to experience significant rapid urbanization, there is an increasing need to access and visualize data which represents the dimensions of space and place (see Straumann et al. (2014) for a distinction of the terms space and place). On this note, the rise of *smart cities* has led to the instrumentation of cities with more real-time data and historical data being captured and visualized (Cheshire & Batty 2012). Large-scale projects such as the Urban Big Data Centre (UBDC) (http://ubdc.ac.uk/) and the Australian Urban Research Infrastructure Network (AURIN) (http://aurin.org.au/) are endeavoring to develop "big data" visualization tools and techniques to support the realization of smart cities. The challenge is to be able to provide not only such data visualization interfaces to researchers, but also tools to support policy and decision makers, city planners and communities to be able to visually explore and analyze this data to make better decisions in collectively planning our cities.

For example, addressing a rapidly urbanizing Australia, AURIN has developed a geoportal where over 1,800 datasets can be accessed and visualized. AURIN has deployed a federated data architecture which is metadata driven (Sinnott et al. 2015). There are over 6 billion data elements that urban researchers, government policy and decision makers can access via the AURIN portal (Pettit et al. 2014). This rich tapestry of "big data" across the domains of health, housing, transport, demographics, economics and other essential areas provides coverage across the major cities of Australia. Both *point in time* and *longitudinal data* are accessible via the AURIN portal. Other examples show how geospatial big data is being used in practice in connection with multi-user (collaborative) visualization environments in the context of geoportals for visualizing big data through work being undertaken in Europe through INSPIRE and in Australia through AURIN.

Another "cutting edge" research area is the visualization of networks, such as transportation networks (Cheshire & Batty 2012). In the London Oyster card case mentioned before in Section 2, there is a need to be able to visualize individual and aggregated travel journeys to provide further insights in the travel behaviors of commuters as they move through the city, and this, in turn, provides important information to transport planners in optimizing timetabling and responding to events. Figure 1 illustrates a snapshot of a time sequence animation of a typical weekday travel based on the Oyster card data.

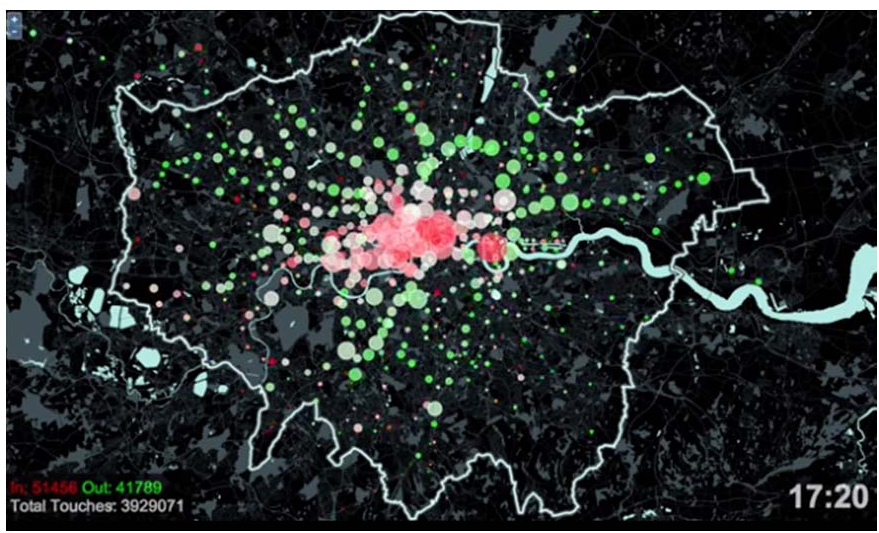

Figure 1 Oyster Card data for London Tube and train stations, animated for a day using 10 minute intervals (Created by Oliver O'Brien (http://oobrien.com/2013/03/londons-tidal-oyster-card-flow/) (via Batty 2012).

In addition, the real-time visualization of crowd-sourced big data from platforms such as Twitter can assist with, for example, disaster management responses. The visualization of real-time data streaming from these sources can provide emergency response team critical information on how to respond to events such as flood, fires and other natural and human induced disasters. Visualization platforms such as Ushahidi (http://www.ushahidi.com/) and Cognicity (http://cognicity.info/cognicity/) and the Peta Jakarta project (http://petajakarta.org/banjir/en/) are such crowd-sourced platforms. In the latter, citizens can tweet the reporting of floods particularly in the Monsoon season in the city of Jakarta, Indonesia.

## 6. Data Mining and Knowledge Discovery

General big data analysis and traditional spatial data analysis and geo-processing methods and theories can all contribute to the development of geospatial big data analysis and processing. Statistical analysis, geo-computing, simulation and data mining methods and techniques can be used alone or together with other types of big data for discovering knowledge from geospatial big data. This section examines traditional knowledge discovery methods and resurgence of fractal analysis in dealing with geospatial big data.

## 6.1 Data mining and knowledge discovery

Knowledge discovery (KD) is concerned with mining and extracting meaningful patterns and relationships from large datasets that are valid, novel, useful and understandable (Miller & Hanz 2009). The field emerged in response to the need for methods applicable to data that violate the assumptions of traditional statistics. KD methods typically emphasise generalisation ability and predictive performance, which is particularly pertinent with spatio-temporal data because spatio-temporal datasets can provide rich information about how a process evolves over time. KD from spatio-temporal data enables us to create models that are able to predict future states of a process, so called the holy grail of science (Cressie & Wikle 2011). KD encompasses a range of spatio-temporal data mining (STDM) tools and methodologies for carrying out a set of tasks.

Perhaps the most conceptually simple KD technique is *Association rule mining* (ARM), which involves searching for associations in a dataset where an event X tends to lead to an event Y, where X is the antecedent and Y is the consequent (Agrawal et al. 1993). In the context of spatio-temporal data, ARM entails searching for the occurrence of an event Y in the spatio-temporal neighbourhood of another event X (Mennis & Liu 2005). Shekhar et al. (2011) describe some of the types of patterns that may be present in spatio-temporal data.

More traditional data analysis methods also come under the remit of KD. *Regression* is viewed as a data mining technique, but in truth its use in the spatial sciences and time series analysis predates the field of data mining. Regression models for spatio-temporal data emerged from the cross-pollination of ideas from time series analysis, econometrics and the spatial sciences. For example, the space-time autoregressive integrated moving average (STARIMA) model (Pfeifer & Deutsch 1980, Cheng et al. 2014), spatial panel data models (Elhorst 2003), Bayesian hierarchical models (Cressie & Wikle 2011) and space-time geostatistics (Heuvelink & Griffith 2010), amongst others. An important form of regression is classification, in which the outputs are class labels. As larger and more granular datasets have become available in recent years, the limitations of traditional statistical approaches in capturing nonlinearity and heterogeneity have been exposed. Therefore, some scholars have looked to the machine learning (ML) community for alternatives, first to artificial neural networks (ANNs) in the 1990s and more recently to kernel methods. Perhaps the most well-known and broadly successful machine learning (ML) method for classification and regression is the support vector machine (SVM), which uses kernels to carry out nonlinear regression or classification (Kanevski et al. 2009, Haworth et al. 2014). Recently, the Random Forest (RF), a method that combines multiple decision trees through bootstrapping, has also gained popularity for classification (Cutler et al. 2007).

Other KD tasks include *anomaly detection*, also known as outlier detection, and *clustering*. Anomaly detection involves the identification of events or patterns in data different from what one would expect. Anomaly detection is inherently challenging as it requires a clear definition of what is considered to be normal and abnormal. In spatio-temporal processes, these definitions may evolve and change over time (Chandola et al. 2009). Accounting for these changes in different spatio-temporal processes is a key research challenge. *Clustering* is a form of unsupervised learning, which involves uncovering hidden structure in a dataset about which we know little. Clustering has wide applications in the spatial sciences, for example in geodemographic classification (Vickers & Rees 2007) and hotspot detection (Nakaya & Yano 2010). Although spatial clustering methods are well developed, spatio-temporal clustering (STC) is still an emerging research frontier. STC methods that are gaining popularity include ST-DBSCAN (Birant & Kut 2007) and space-time scan statistics (STSS) (Kulldorff et al. 2005, Cheng & Adepeju 2014).

The impact of ML methods on KD and STDM has been significant. ML methods are generally effective in tackling nonlinearity in spatial data, and can be modified to deal with the multi-scale issue and heterogeneity (Foresti et al. 2011). However, in many cases (especially kernel methods) their initial computation is expensive if the number of data samples is large. Furthermore, if the statistical properties of a space-time series evolve over time, models have to be retrained to reflect this. In the big data age where the ability to apply methods to real time data streams is paramount, new ways of training traditional algorithms are needed. In ML, online learning is used (Castro-Neto et al. 2009) and these types of approaches need to be integrated with more traditional spatio-temporal analysis techniques. Parallel and grid computation can also be used to improve the performance of KD methods (Harris et al. 2010). However, there are issues that cannot be solved with improvements in computational efficiency alone. For example, the principal problem in STC is to model how clusters emerge, change, move and dissipate/disappear in time. This can be achieved retrospectively but is very difficult to quantify in time critical applications. At what point does a cluster of crimes become a hotspot? Current methodologies within KD and STDM are generally designed to analyse historical datasets but cannot adequately deal with evolving properties of space-time data.

## *6.2 Fractals emerged from big data*

Big data show incredible fractal structure, in which there are far more small things than large ones. There are several reasons for the emerging fractals. First, big data usually emerge from the bottom up or are contributed by diverse individuals, e.g., location based social media data, so they are very diverse and heterogeneous. Second, big data are defined at very high spatial and temporal scales, which enable us to observe fractal structure and nonlinear dynamics more easily. Third, big data due to the size are more likely to capture a true picture of reality, and reality is no doubt fractal (Bak 1996, Mandelbrot & Hudson 2004). In these aspects, the emerging fractals differ from fractals seen on small data such as fractal cities (Batty & Longley 1994). We argue that fractal geometry (Mandelbrot 1982), or complexity science in general, should be adopted for big data analytics and visualization, and for developing new insights into big data.

Let us examine a working example to see how fractals are generated from tweets location data of the entire world (Jiang 2015a, Jiang & Miao 2015). The data were sliced at different intervals, and in an accumulative manner, which means that locations at t1 are included in locations at t2. For each sliced location dataset, we built a triangulated irregular network (TIN), and merged those small TIN edges (smaller than the mean of all the TIN edges) as individual patches, which are referred as natural cities. Eventually, thousands of the natural cities are emerged from the tweets locations. Figure 2 shows two natural cities near Chicago and New York at the four time instants, and they are put in comparison with the generative fractal – Koch snowflake. The two natural cities look very much like the snowflake, both sharing (1) the scaling pattern of far more small things than large ones, and (2) irregular shapes or boundaries. On the other hand, the natural cities look a bit different from the snowflake; the natural cities are developed from some irregular patches, which become further fragmented, whereas the snowflake and its growth come from the regular triangle with a strict scaling ratio 1/3. In other words, the former is called statistical fractal, being statistical self-similar, while the latter called strict fractal, being strict self-similar.

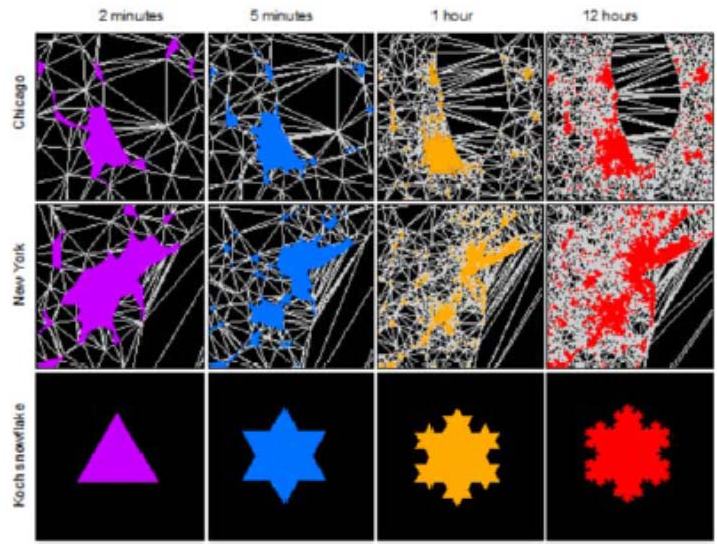

Figure 2 Fractals emerged from big data look very much like the generative fractal – Koch snowflake

What the example illustrated are not only fractals or natural cities generated from big data, but also a new, relaxed definition of fractals. A set or pattern is fractal if there are far more small things than large ones in it, or the scaling pattern of far more small things than large ones recurs multiple times (Jiang 2015b, Jiang & Yin 2014). The new definition is in fact developed from head/tail breaks (Jiang 2013) as a classification scheme for data with a heavy tailed distribution. The complexity of fractals can be measured by the head/tail breaks induced index – ht-index: the higher the ht-index, the more complex the fractal. Conventionally, complexity was captured by fractal dimension (Mandelbrot 1982), thus ht-index being an alternative index to fractal dimension. In comparison to conventional definitions of fractal, the new definition is much more intuitive and easier to understand, so that anyone can rely on it to see fractals. For example, society is fractal because there are far more poor people than rich people, or far more ordinary people than extraordinary people (Zipf 1949). It should be noted that the head/tail breaks method is not only for data classification, with which both the number of classes and class intervals are automatically determined, but also an efficient and effective visualization tool (Jiang 2015b). Figure 3 presents an example of visualization using the head/tail breaks; note that only part of the whole is shown to the right panel, but the part is self-similar to the whole, and thus the part reflects the whole.

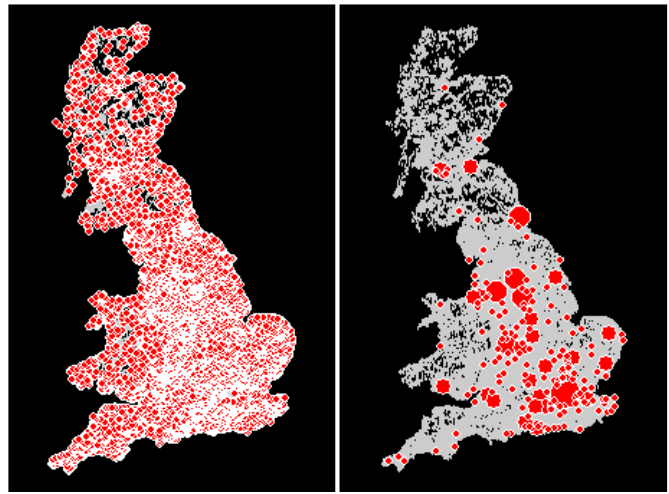

Figure 3 Head/tail breaks as an effective visualization tool (Note: 16,000 natural cities (to the left ) in UK generated from a half million of points of interest look like a hairball, but their top 4 hierarchical levels clearly show a scaling structure to the right.)

Fractal geometry represents a new way of thinking for geospatial analysis, and this is particularly truly for big data analytics. Despite that Euclidean geometry has thousands of years of history, and serves as the foundation of geospatial technologies, the essence of geography (both physical and human) is fractal. We therefore must adopt fractal methods for developing new insights into big data. This fractal thinking is in line with Paretian thinking (Jiang 2015b), which is in contrast to conventional Gaussian thinking. Statistically speaking, there are far more small things than large ones, rather than more or less similar things. Given the scaling pattern or the heavy tailed distribution, the head/tail breaks provide an effective means to derive an inherent hierarchy of complex systems. This is the first step towards an understanding of geographic phenomena, i.e., to recognize fractal structure of geographic systems. The next step is to further develop, through simulations, an understanding of processes as to why the fractal structures exist.. In this regard, a set of complexity modeling tools have been developed such as cellular automata, agent-based modeling, and sand pile model.

## 7. Challenges, Open Issues and Future Directions

This section presents the challenges and open issues based on the reviews in Sections 2-6 and outlines some research directions in the three focus areas of ISPRS TC II (http://www2.isprs.org/commissions/comm2.html).

### *7.1 Efficient representation and modelling for geospatial big data*

The main problem that spatial algorithms face in this context is that they cannot wait until all the data are known, as it is the case in two of the major classes of spatial algorithms: the divide-and-conquer or line/plane-sweep algorithms. Even though incremental algorithms are well-suited for handling changing data sets, they are not well-suited for handling streams that cannot fit in the main memory of the computer. Therefore, a new class of spatial algorithms has started to be designed and developed: streaming spatial algorithms. However, any real-time streaming algorithms must read an input stream, process it, and write the output stream in real-time. Among all the tasks that have to be performed on computers or digital circuits, the tasks that are the slowest, i.e., communication over a network, must be minimized in order to satisfy the real-time constraint of big data processing. Another implication of the constraint of real-time processing of big spatial data is to minimize the amount of disk input/output, which is the second slowest task after communications over a network. For this purpose, one needs to focus on the most compact data structures that will store spatial data using the smallest amount of memory. Finally, as spatial data tend to be more complex than non-spatial data, the total CPU running time is not negligible with respect to the disk input/output time. Thus, the parallelization of the spatial algorithms will bring a non-negligible speed-up to the spatial algorithm. This means the spatial algorithms must be distributed, parallelized and use the most compact spatial data structures so that the exchange of information over a network and between the disk and the main memory are minimized and the CPU running time is also minimized. This imposes a distributed parallel architecture where streams of data are processed in real time on each unit controlling a sensor in order to transmit over the network only summary statistics and required results for the other nodes in the computing environment. Such summary statistics that are relevant for sensors are intervals, and in particular the new measurements that change the intervals of measurements or the required results.

Furthermore, interval analysis is in conjunction with functional analysis, the most-well-suited framework to model the uncertainty of geospatial big data. Firstly, as it has been already observed by Juditsky et al. (1995), functional analysis methods like wavelets do not suffer from the curse of dimensionality that affects machine-learning as well as parametric statistics (including multivariate statistics). Secondly, interval analysis allows one to model the uncertainty of the input variables (like sensor observations) and the corresponding uncertainty of the functions that are evaluated on these variables. The main challenge and open issue is to bridge the gap between machine learning and functional analysis communities by convincing the communities working on and with geospatial big data to use functional analysis and interval analysis with a functional programming language. Unfortunately, even though functional programming has made its way into other programming language paradigms, functional analysis combined with interval analysis has not been so successful in making its way in applications of big spatial data.

Four initial observations can be seen as the base for exploring further research directions. Firstly, functional analysis methods like wavelets, homotopy continuation and interval analysis are much better suited than parametric statistical methods to cope with the curse of dimensionality inherent in big spatial data that includes many dimensions (each functionally independent physical value measured corresponds to one dimension and each one of the coordinate systems component x, y and z corresponds also to one dimension). Secondly, pure functional programming is very well suited for handling functions because pure functional programming does not have (unlike impure functional programming and other programming paradigms) side effects in pure functions, and functions are one of the two most fundamental concepts in any functional programming language: functions and data types. Thirdly, the main challenge in spatial data handling, which is to certify the topological relationships and the uncertainty of any spatial data modelling or decision-making by determining the uncertainties of all input variables, can be solved using interval analysis. Finally, the lazy functional programming paradigm is very well suited for fractals due to its ability to represent fractal recurrences and to compute only the fractal components that are needed to compute the final result.

One future research focus in his area is to produce a locally distributed stream sensing, processing and telecommunicating paradigm implemented using:
- new functional specification methods derived from ontologies, ontology mappings and their gluing into ontology categories and their charts;
- functional analysis methods (decompositions of streams with interval enclosures of wavelets, mathematical modelling in higher dimensional measure spaces from interval valued homotopies in lower dimensional measure spaces);
- pure functional programming languages (see (haskell.org committee, 2015)) with:

- visualization libraries (like Haskell and its OpenGL bindings and many specialized visualization libraries),
- parallelization and concurrency libraries (like the distributed MapReduce framework Holumbus or OpenCL and OpenMP wrappers),
- cloud computing libraries;
- functional programming inspired parallelisation and concurrency techniques.

The architecture of any system based on this paradigm can be considered a fractal. Every sensor controlling unit is responsible for collecting the big data streams, computing the statistics or any other desired result, generating the triggers that will automatically update any computed result, visualization or decision making and transmitting it to a lower-resolution data collecting node. Each local data-collecting node is responsible for storing the streams of results and providing the desired visualization and assembling the partial decision taking elements into a summarized decision taking from the neighbouring sensor controlling units. Each regional data collecting node is responsible for storing the streams of results and providing the desired visualization and assembling the partial decision taking elements into a summarized decision taking from the neighbouring local data collecting nodes.

The other future research focus on new processing algorithms to handle large volumes of data through use of functional programming languages is to design new streaming algorithms that:
- use the provably most compact geometric topology data structure that encodes all the ramifications (i.e. the singular points of the skeletons of the objects),
- use CPU and GPU parallelization to harness the computing facilities wherever these lie (locally),
- use interval analysis to:
  - represent uncertainty of streams of spatial data (Dilo et al. 2007), and
  - to automatically generate the triggers that will react automatically once an input value in a stream will make a change in any computed result, visualization or decision making;
- use wavelet decompositions for handling signals acquired by sensors,
- use fractals for handling spatial data whose Hausdorff-Besicowitz dimension is not an integer,
- use interval valued homotopies to model or reconstruct functions in higher dimensional measure spaces from measurements and reconstructions of functions from lower dimensional measure spaces, reconstructing therefore higher dimensional measure spaces one dimension at a time,
- use categories to represent the functional dependencies between data variables and any computed result or visualization or decision taking (like data types and functions in the Hask category), and a category based Domain Specific Language computing library (like docon, see Mechveliani (2006)).

## *7.2 Analysing, mining and visualising geospatial big data for decision-support*

Spatial statistical methods are in principle able to also include non-spatial big data. A typical example is the use of co-variables in a spatial interpolation procedure (Van de Kassteele & Stein 2006), or in a spatial point process modelling analysis. Such analyses commonly rely on assumptions, such as normality, independence, absence of noise in the explanatory data. With the advancements of big data, however, on the one hand the quality of the big data can be doubted, whereas on the other hand the speed of calculations is seriously affected. Hence, the procedures are at the moment not ready for the purpose, and serious pre-processing has to be done, for which the tools are not readily available. Some of the methods have been developed in the past, such as possibilities to use explanatory variables of a changing quality, but these are often rather complicated to use, and require substantial work to make them available for the purpose. As an example one may think of Bayesian procedures, where prior distributions can be included into a likelihood function. These methods have shown an enormous power, also in spatial studies; their application on an automatic basis for big data may not be so easy and transparent to develop, implement and apply.

The main challenges in KD and STDM can be neatly summarised around the first four *Vs* described in Section 1.

*Volumes* of spatio-temporal data are ever increasing and some argue that traditional transactional database structures are becoming outmoded. Although this point is open to debate, research is taking place into how to best use new data storage and query architectures to deal with spatio-temporal data. For example, recent studies have used MapReduce programming models to parallelise data processing (Tan et al. 2012). Despite this, traditional SQL based spatial databases, such as PostGIS and Oracle Spatial, remain dominant in academia. If this trend continues, there is a risk that academic research will become disconnected from industry.

Despite significant progress, challenges remain in developing predictive algorithms that can deal with the *velocity* of data arriving in real time. Geography has been a big data discipline since long before the term arose, dealing with large and complex problems like weather and climate modelling. Hence, geography has traditionally been forward thinking regarding the development of algorithms for dealing with large and complex datasets in a timely fashion. The discourse on parallel computing in geography began in the 1990s, leading to the emergence of the subfield of Geocomputation (Cheng et al. 2012). However, parallelism is still far from the common practice despite enabling hardware being available in desktop computers. Most recent work has focussed on parallelising all or part of existing algorithms to improve computational performance. For example, Guan and Clarke (2010) developed a parallel raster processing library for use in cellular automata, and Guan et al. (2011) developed algorithms to parallelise elements of Kriging interpolation. Libraries for parallel geocomputation are also now beginning to emerge in open source software environments such as R (Harris et al. 2010). Building on this, a greater focus needs to be placed on developing algorithms that are parallel in nature and can harness all types of parallelism. This is what Turton and Openshaw (1998, p. 1842) termed "Thinking in Parallel" in 1998, but is yet to be fully adopted in the research community.

*Variety* presents itself as an opportunity to the research community. We now have potentially many data sources of different types that can be used to analyse (and re-analyse) a diverse range of spatio-temporal processes. Traditional data providers such as governments, national mapping agencies and transport authorities are now complemented by new data sources described in Section 2. However, one of the important and often overlooked problems of KD in the spatial sciences is the issue of data *veracity*. Big data gives us unprecedented volumes of data pertaining to a broad range of human activities and physical processes. However, they are often collected on a fairly ad-hoc basis when compared with traditional data sources, and usually must be repurposed to fulfil research objectives. Initiatives such as OpenStreetMap have proven that crowd sourced data can compete with official sources in this regard, but issues of sample representativeness and data collection design (or lack thereof) are still a concern. A good example is Twitter, which has gained popularity in research communities recently, but under represents certain groups, including the elderly and some ethnic minorities, and has an uneven spatio-temporal distribution. Such datasets clearly have potential value alongside traditional demographic data such as censuses and cross-sectional surveys (Longley et al. 2015). However, there is a trade-off between the spatial and temporal granularity offered by these data and the certainty associated with any conclusions drawn from them. Personal mobility data is another example; smartphone apps collect and store vast quantities of such data which have considerable research potential, but this cannot be fully realised without proper validation and a clear understanding of potential biases in the user group. This is coupled with difficulties in inferring activities from raw, unlabelled mobility data (Bolbol et al. 2011).

Sampling is not a legitimate concept in the big data era, as argued by Jiang and Miao (2015). Big data tends to take all or a large amount, and this data characteristic make big data different from small data, which are often sampled. Surely, social media data are oriented towards younger generations or those who have access to Internet and social media, and not everyone has Internet access, in particular in developing countries. However, the large data volumes enable big data to capture the true picture of all, despite the fact that not all people are involved in social media. Big data calls for new ways of thinking, while our thinking tends to be conventional (Jiang 2015b, Mayer-Schönberger & Cukier 2013). For example, we tend to examine data quality as we did in the small data era. Massive data should tolerate messiness. For navigation purposes, we must make OSM data as precise and accurate as possible, in other words, the more precise and the more accurate, the better. However, to examine whether there are far more small street blocks than large ones, we do not need very good quality data, but we need massive data with messiness instead.

Nonetheless, visualizing big data has some additional challenges; for example, visual analytics solutions themselves may not scale: we need to consider how to deal with *information overload* on the viewer if we show too many things at the same time (Choo & Park 2013, Ruff 2002). Applications in a large-scale geoportal like AURIN have revealed there are significant challenges in being able to visualize large multi-dimensional geospatial datasets via a browser. For this, thousands of years of cartographic expertise offers sound advice. We should carefully generalize, e.g., emphasize the important while removing the unimportant, group the information both thematically and perceptually, and pay attention to visual hierarchy when we design displays.

We should remember linking the good cartographic design principles to modern interaction design paradigms (MacEachren et al. 2008, Schnürer et al. 2014, Slocum et al. 2008). Furthermore, researchers in the cartography and geovisualisation domain have taken a strong interest in cognitive and usability issues and much progress has been made to understand how human capacity can enhance or limit our experiences with visual displays (Çöltekin et al. 2010, Knapp 1995, Montello 2002, Roth 2013, Slocum et al. 2001).

Whilst the *smart city* and *big data* are hot topics of today (Batty 2012), there is also the challenge of how to visualize the error and uncertainty inherent with big data sets such as crowd-sourced datasets and smart card data (Cheshire & Batty 2012). In the realm of geospatial big data, there is also a need to effectively visualize other types of massive data sets, such as LIDAR data or very large collections of remote-sensing data. Visualization platforms such as PointCloudViz (http://www.pointcloudviz.com/), or Online LIDAR point cloud viewer (http://lidarview.com/) specialize on Lidar data visualization. Similarly, there are efforts in organizing remotely sensed imagery (Ma et al. 2014, Marshall & Boshuizen 2013).

Spatial statistics has always had good opportunities for visualizing spatial and spatio-temporal data, including the uncertainties (see Rulinda et al. (2013) for an excellent example in the space-time domain). Their opportunities are still present also for geospatial big data. In the past, such methods have also been applied to non-spatial data, and there is apparently not a real problem to extend them towards big data. A critical issue in all of spatial (or spatio-temporal) data is their relying on coordinates. Hence, also for non-spatial data there must be an opportunity to assign coordinates, or equivalent, to the non-spatial data. Successes in the past have typically considered class memberships to serve that purpose, bringing the non-spatial data towards the feature space.

## *7.3 Quality assessment of geospatial big data from new sources*

Dealing with veracity in a scalable and timely manner has been identified as a substantial challenge (Saha & Srivastava 2014). Behind this challenge is the lack of thorough knowledge of the data semantics when data that is commonly called "dirty" (e.g., Chiang & Miller 2008) is collected and combined in an ad-hoc manner. Accordingly, big data analytics has given up the closed-world assumption in favour of learning incrementally under an open-world assumption. Thus, big data quality assessment has been characterized by the three stages of discovering rules of data semantics, checking for inconsistencies based on currently known rules, and repairing data near real-time (Saha & Srivastava 2014).

Geospatial big data collection, especially sensors' data, is being captured in an automatic and unprecedented way, which poses new opportunities and challenges. This in principle causes problems for a statistical analysis, where statistical considerations such as optimal design or model based sampling are critical to make valid statements. Big spatial data may be either largely irrelevant, as they are collected automatically and hence much uninteresting repetition may occur, whereas big spatial data may be of a highly varying quality. They may be precise, but too precise to be of use, they may be too abundant as the same object is over sampled, they may be of a very poor quality as data may be at the nominal scale, where ratio-scale data might have been collected elsewhere, or they may be inadequate for the specific purpose. At the moment, no adequate procedures seem to be implemented to be able to harmonize the quality of spatial data for specific purposes. A big step has to be made to be able to overcome this.

On the positive side, the large number of data sources allows us to acquire a "complete picture" of a spatial situation, also including its dynamics. This is even more so, when different sensor data are integrated and fused, allowing to also eliciting more than one aspect or feature of an object (Ebert et al. 2009). Due to the potential high redundancy, it is possible to identify errors or blunders in the data and achieve higher accuracies even though an individual sensor has a limited measuring quality.

As often no explicit semantics (or manual annotation of semantics) is given, automatic processes are required to reveal it. This refers both to the object and its features, but also to the temporal characteristics.

Crowd-sourced data sets are often related to very detailed, i.e., large-scale phenomena. Still, however, the level of granularity (in space, time and semantics) is not necessarily known. A challenge is therefore, to determine this granularity level from the mere sensor data. One option is to include explicit metadata about the sensor in terms of a self-description. Another option is to infer the scale and granularity level by automatically relating it to other sources of known granularity. This involved matching techniques on not only a geometric but also a semantic level.

In traditional sensors the quality (e.g., geometric accuracy) of the acquired data is given. This is not necessarily the case with new data sources such as VGI data. As this data source is often user centred, questions of reputation and trust have to be included to evaluate the quality.

## 8. Conclusions

This study reviews a variety of geospatial theory and methods that have been used for traditional data but that can be extended to handle geospatial big data. While there is no standard definition of big data, it can be considered as structured and unstructured datasets with massive data volumes that cannot be easily captured, stored, manipulated, analysed, managed and presented by traditional hardware, software and database technologies. Given these unique characteristics, traditional data handling approaches and methods are inadequate and the following areas were identified as in need for further development and research in the discipline:

- The development of new spatial indexing and algorithms to handle real-time, streaming data and to support topology for real-time analytics.
- The development of conceptual and methodological approaches to move big data from descriptive and correlation research and applications to ones that explore casual and explanatory relationships.
- The development of efficient methods to display data integrated in the three dimensions of geographic and one dimension of continuous time. There is a strong need in understanding human capacity to deal with visual information and identifying which visualization type is a good fit for the task at hand, and the target user group. Furthermore, interdisciplinary studies and communication is of critical importance. The advances in scientific visualization and information visualization are both beneficial to geographic visualization; but geographic visualization has also a lot to offer to other domains. Novel visualization paradigms, especially developed for big data tend to be information-rich (thus complex); therefore, we find that highlighting and summarizing approaches should be further investigated. Additionally, and in relation to managing complex visualization displays, technology research in terms of level of detail management remains important.
- The development of novel approaches for error propagation so as to effectively assess data quality requires. The challenge is not only the handling the many different types of data for real-time analytics, but rather the ad-hoc combination of data streams in real-time, which may include the capture of the "whole" picture (or "complete population") instead of sampling a small portion of the whole population. In this case quick assessments are preferable that may come out of varying the input data and simulating variability.

In addition, other general conceptual and practical issues were also highlighted. The relation between spatial statistics and semantics and ontologies has been identified in the past, but deserves to be further elaborated. Ontologies have been identified, for example, for dunes and beaches in studies around 2005, whereas more slum ontologies have been developed. The role of spatial statistics was related to the scale, the environment and the characterization of specific variables. In particular aspects of scale are important and have shown their relevance.

Privacy and security are equally important and key concerns especially in geospatial big data handling, which may lead us into a "naked future" if not properly addressed. They are an essential part of geospatial big data management, but are not covered here given the focus on this paper on data handling methods.

Big data presents both challenges and opportunities. This paper outlines some of these challenges from a technical and conceptual perspective, and also provides priority areas that need to be addressed in the future. Once big data research matures, the opportunities for overall societal management and decision-making become enormous.

## Acknowledgment

The authors would like to thank Anthony Lee (Simon Fraser University) for his help in compiling all the references and formatting citations.

## Authors Contribution Statement

Regardless of the order of authors listed, all authors contribute equally to the paper by participating discussions, writing sections, revising corresponding sections and providing revision comments on the whole manuscript.